\documentclass[conference]{IEEEtran}
\IEEEoverridecommandlockouts
\usepackage{cite}
\usepackage{amsmath,amssymb,amsfonts}
\usepackage{algorithmic}
\usepackage{graphicx}
\usepackage{textcomp}
\usepackage{xcolor}
\def\BibTeX{{\rm B\kern-.05em{\sc i\kern-.025em b}\kern-.08em
    T\kern-.1667em\lower.7ex\hbox{E}\kern-.125emX}}
\begin{document}

\title{Agricultural 4.0 Leveraging on Technological Solutions: Study for Smart Farming Sector%\\
%{%\footnotesize \textsuperscript{*}Note: Sub-titles are not captured in Xplore and

}

\author{\IEEEauthorblockN{Emmanuel Kojo Gyamfi, Zag ElSayed, Jess Kropczynski, Mustapha Awinsongya Yakubu, Nelly Elsayed}
\IEEEauthorblockA{\textit{School of information Technology} \\
\textit{University of Cincinnati}\\
Ohio, United States \\
%gyamfieo@mail.uc.edu, elsayezs@ucmail.uc.edu, jess.kropczynski@uc.edu, yakubuma@mail.uc.edu, elsayeny@ucmail.uc.edu
}
}

\maketitle

\begin{abstract}
By 2050, it is predicted that there will be 9 billion people on the planet, which will call for more production, lower costs, and the preservation of natural resources. It is anticipated that atypical occurrences and climate change will pose severe risks to agricultural output. It follows that a 70\% or more significant rise in food output is anticipated. Smart farming, often known as agriculture 4.0, is a tech-driven revolution in agriculture with the goal of raising industry production and efficiency. Four primary trends are responsible for it: food waste, climate change, population shifts, and resource scarcity. The agriculture industry is changing as a result of the adoption of emerging technologies. Using cutting-edge technology like IoT, AI, and other sensors, smart farming transforms traditional production methods and international agricultural policies. The objective is to establish a value chain that is optimized to facilitate enhanced monitoring and decreased labor expenses. The agricultural sector has seen tremendous transformation as a result of the fourth industrial revolution, which has combined traditional farming methods with cutting-edge technology to increase productivity, sustainability, and efficiency. To effectively utilize the potential of technology gadgets in the agriculture sector, collaboration between governments, private sector entities, and other stakeholders is necessary. This paper covers Agriculture 4.0, looks at its possible benefits and drawbacks of the implementation methodologies, compatibility, reliability, and investigates the several digital tools that are being utilized to change the agriculture industry and how to mitigate the challenges.
\end{abstract}


\section{Introduction}

The term "Agriculture 4.0" refers to the industry's upcoming significant trends, such as the internet of things (IoT), using big data and Machine learning to make businesses more efficient amid the challenges of population growth and climate change to improve output, efficiency, and support sustainable agriculture by using accurate information to make strategic decisions~\cite{ref1}.

\begin{figure}[t]
	\centerline{\includegraphics[width=8.5cm, height= 4 cm]{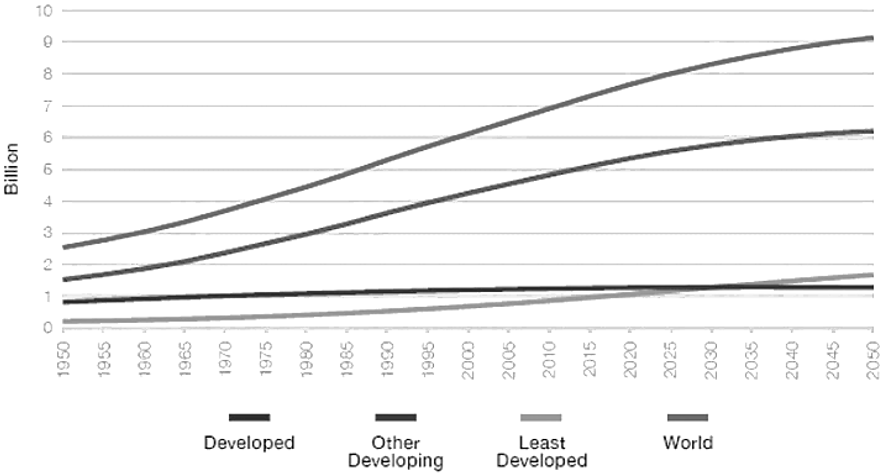}}
	\caption{The statistics of the world population growth according to the Food and Agriculture Organization (FAO) 2009, Global agriculture towards 2050.}
	\label{fig1}
\end{figure}

The agricultural sector, widely recognized as a fundamental pillar of human civilization, has undergone a persistent process of evolution throughout history~\cite{ref2}. Since the advent of agriculture more than ten millennia ago until the green revolution of the 20th century, the human pursuit of cultivating crops and raising livestock has transformed to meet the increasing needs of a rapidly expanding global populace. Figure~\ref{fig1} shows the of the world population growth accorfing to the Food and Agriculture Organization (FAO) 2009, Global agriculture towards 2050. As we approach the advent of the fourth industrial revolution, the agricultural sector finds itself on the verge of experiencing another significant paradigm shift~\cite{ref3}. This forthcoming transformation is expected to be driven by advancements in technology and the seamless integration of digital systems. This change possesses the capacity to fundamentally alter food production, bolster sustainability, and tackle the complex array of difficulties that afflict our worldwide food system.

The concept of Agriculture 4.0 is undoubtedly a promising technology. In 2018, the World Government Summit (WGS) collaborated with Oliver Wyman to release a research paper on this good technology that seeks to revolutionize our future. In this paper, the authors mention four primary trends that can put pressure on agriculture soon. The trends include changes in climate, wastage of food, demography, and scarcity of natural resources~\cite{ref4,ref5}. 

The implications of Agricultural 4.0 are far-reaching and multifaceted. Farmers, who are responsible for managing the world's cultivable land, can achieve benefits through increasing crop production, minimizing input costs, and adopting sustainable methods in agriculture. Concurrently, stakeholders across the agricultural value chain, such as agribusinesses, governmental bodies, researchers, and consumers, will experience notable changes in supply chain transparency, food quality, and environmental stewardship. 

Nevertheless, this process of technological transition is not without its associated obstacles. Relevant issues encompass concerns over data security and privacy, the potential exacerbation of inequities through the digital divide, and ethical dilemmas associated with automation and genetic engineering~\cite{ref6}. Furthermore, it is imperative to recognize that the trajectory of agriculture is inextricably linked to the far-reaching consequences of climate change, hence requiring the implementation of agricultural methods and policies that are resilient to climate variations and capable of adapting to changing conditions.

Ara{\'u}jo et al.~\cite{ref2} highlight the potential of emerging technologies in agriculture, such as machine learning, big data analytics, artificial intelligence, the Internet of Things, and drones, in increasing revenue and input. However, these technologies will not be distributed uniformly, and careful discussion is needed to ensure their full potential benefits are spread across countries. Artificial intelligence automates tasks and makes decisions based on data, while IoT devices collect and exchange data. Drones are used for surveillance and mapping, and gene editing allows scientists to modify living organisms' genetic code.

The study explores Agricultural 4.0's origins, technological foundations, implementations, advantages, obstacles, and future developments. It aims to understand how it reshapes agricultural methodologies, revitalizes rural societies, and addresses global food access, sustainability, and climate change resilience. Smart farming, or Agricultural 4.0, is the future of agriculture, shaped by embracing technologies and recognizing their potential to create abundant food, thriving ecosystems, and sustainable farming practices.

\section{Literature Background}
Agriculture has long served as the fundamental pillar of human civilization, offering nutrition and livelihoods for many generations. Over the course of history, agricultural practices have undergone transformations to adapt to shifting natural circumstances and the increasing needs of a worldwide populace. The shift from subsistence farming to industrial agriculture throughout the 19th and 20th centuries marked notable progress in the areas of mechanization, crop breeding, and chemical inputs. These innovations resulted in heightened production and enhanced food security~\cite{ref7}. Nevertheless, the agricultural industry has encountered a fresh array of obstacles in the 21$^\mathrm{st}$ century. The issues encompassed in this context consist of climate change, limited availability of resources, population expansion, and the imperative for the adoption of sustainable farming methods. Considering these issues, a paradigm change referred to as Agricultural 4.0, or Smart Agriculture, has arisen.

To meet the demands of a projected world population estimated at 9 billion by 2050. There is a need to enhance productivity, minimize costs, and adhere to the principles of protecting our natural resources. In addition, it is anticipated that climate change and the occurrence of unusual occurrences will pose a significant threat to agricultural production. To overcome these challenges, emerging technologies must be adopted in agriculture. These technologies have the capacity for transformative effects, ultimately leading to the culmination of the fourth agricultural revolution~\cite{ref9}.

Agriculture 1.0, an age-old method, predominantly depended on manual effort and animal strength for farming activities. Farmers employed rudimentary implements such as manual plows and sickles, while their irrigation systems were elementary in nature. This traditional farming approach, which depended on manpower and natural resources, lacks contemporary technology or advanced techniques. The inefficiency can be attributed to its dependence on manual labor and the absence of technological advancements that could significantly enhance production~\cite{ref9,ref10}.

The mechanization revolution, taking place in the late 18$^\mathrm{th}$ and 19$^\mathrm{th}$ centuries, brought about a transformation in agriculture, transitioning it from Agriculture 1.0 to Agriculture 2.0. The advent of tractors, steam engines, and mechanical reapers enhanced efficiency, minimized the need for manual labor, and improved crop output, thus fostering food stability and economic expansion. The objective of this revolution was to optimize efficiency in all facets of agriculture, encompassing cultivation, harvesting, and storage, thus diminishing production costs. 

The Green Revolution, also known as Agriculture 3.0, originated in the mid-20$^\mathrm{th}$ century and centered around breakthroughs in crop breeding, the development of high-yielding cultivars, the use of synthetic fertilizers, and the application of pesticides. As a result, crop productivity increased, and food scarcity decreased. Nevertheless, it also elicited environmental and sustainability apprehensions due to heightened chemical inputs. Biotechnological advancements have facilitated the genetic alteration of crops to enhance their productivity and resilience against diseases. Agriculture 3.0 is highly dependent on external inputs, which may result in environmental deterioration. 

\begin{figure}[t]
	\centerline{\includegraphics[width=8.5cm, height= 6 cm]{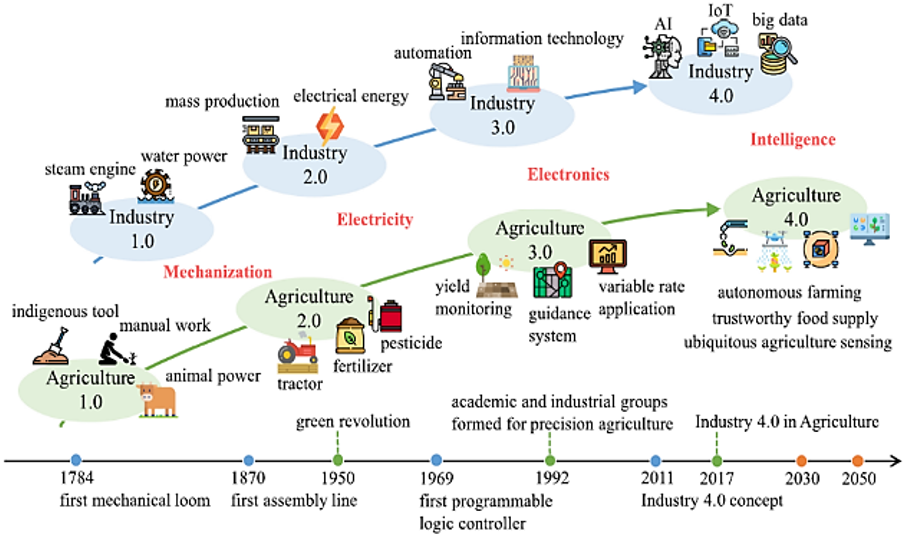}}
	\caption{The development roadmap of the revolution of Agriculture 1.0 to 4.0~\cite{ref10}.}
	\label{fig2}
\end{figure}

Agriculture 4.0 refers to the fourth industrial revolution in agriculture, which involves the integration of sophisticated technologies such as IoT sensors, robotics, artificial intelligence, big data analytics, and cloud computing platforms to digitize and enhance agricultural processes. This revolution aims to optimize agricultural methods, improve efficiency, minimize the environmental impact, and promote sustainable development~\cite{ref2}. It tackles issues related to managing resources, ensuring food safety, facilitating market access, enhancing agricultural productivity, and achieving sustainability objectives by employing intelligent decision-making procedures that rely on real-time data from diverse sources within an agriculture ecosystem~\cite{ref9,ref10}. Figure~\ref{fig2} shows development roadmap of the revolution of Agriculture 1.0 to 4.0~\cite{ref10}

Agriculture 4.0, which can also referred to as The Digital and Data-Driven Revolution, represents the current and ongoing phase of agricultural evolution~\cite{ref11}. It has emerged in response to contemporary challenges such as climate change, resource scarcity, and the need for sustainable agriculture, suggesting that the agricultural production chain involves various stages and processes, including pre-field, in-field, and post-field, focusing on multiple parameters such as soil, environmental conditions, plant and animal characteristics, input application, harvesting, and production~\cite{ref12}. This contributes to various advantages and benefits, including increased productivity, quality, and efficiency, by optimizing the utilization of natural and environmental resources. To achieve the desired impact, emerging technologies, like cloud computing, IoT, big data analytics, machine learning, artificial intelligence, automation, robotics, digital twins, biotechnology, etc., must be adopted. However, Agriculture 4.0 faces numerous interconnected obstacles that hinder the acceptance of these new emerging technologies into the industry~\cite{ref8}.

The notion of Agricultural 4.0 signifies a break from conventional farming practices, which frequently rely on manual labor. In contrast, the focus is placed on precision agriculture, wherein real-time decision-making by farmers is informed by data obtained from sensors and Internet of Things (IoT) devices. An illustration of the potential benefits of soil sensors is their ability to furnish data regarding moisture levels, thus facilitating accurate scheduling of irrigation [13]. Additionally, artificial intelligence algorithms can be employed to scrutinize satellite pictures, thereby enabling the identification of initial indications of crop stress or disease~\cite{ref14}.

The advantages of Agricultural 4.0 are many and varied. According to~\cite{ref3}, using this technology presents the opportunity to augment agricultural productivity, preserve valuable resources, mitigate ecological repercussions, and improve the safety and traceability of food. According to~\cite{ref16}, farmers can enhance their productivity and profitability by allocating resources and minimizing waste effectively. Additionally, these methods align with sustainable agricultural principles.

The implementation of Agricultural 4.0 faces some significant challenges, such as data security, privacy, infrastructure restrictions, and the digital divide~\cite{ref6}. Critically looking at some ethical considerations is also very crucial, as automated farming systems may worsen animal wealth disparity in the agricultural sector. Addressing these issues is essential for fair access and responsible agricultural sector implementation~\cite{ref17}. 

The shift from Agriculture 1.0 to Agriculture 4.0 is a significant transformation, combining advanced technologies and data-driven decision-making to enhance production, promote sustainability, and address key issues like education, accessibility, and responsible deployment in the sector.

\section{Key Technology in Agricultural 4.0}

\begin{figure}[t]
	\centerline{\includegraphics[width=8.5cm, height= 6 cm]{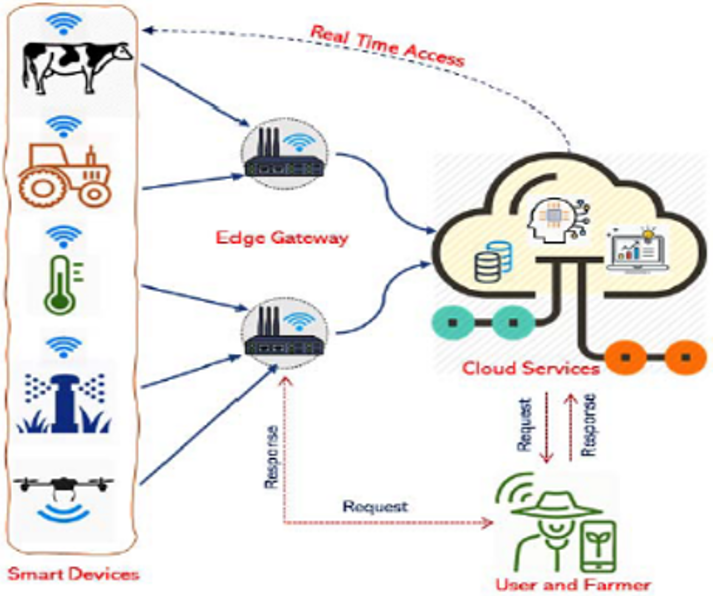}}
	\caption{Smart farming conoceptual architect framework ~\cite{ref10}.}
	\label{fig3}
\end{figure}

Smart Farming signifies a technological paradigm shift within the agricultural industry~\cite{ref7}. Figure~\ref{fig3} shows thefarming conoceptual architect framework~\cite{ref6}. This section delves into the key technologies and roles that have significantly transformed agriculture, focusing on Agricultural 4.0 and their role in driving smart farming.

\subsection{Internet of Things (IoT)}
The Internet of Things refers to a network of interconnected physical devices equipped with sensors, software, and connections that assist the devices in gathering and exchanging data. These IoT devices are primarily used in the agricultural sector to help monitor and exchange data on various platforms, such as soil moisture, determine the temperature and humidity, and observe animal behavior in fields, greenhouses, and livestock facilities and management. The utilization of Internet of Things (IoT) devices facilitates the acquisition of real-time data, augmenting resource management practices' efficiency and fostering improvements in productivity and resource preservation~\cite{ref18}.

\subsection{Big Data Analytics}
Big data analytics incorporates the process of gathering, retaining, and examining extensive and comprehensive databases. Within the realm of agriculture, this technological advancement facilitates the analysis and utilization of data derived from diverse sources, including IoT devices, satellites, weather stations, and historical agricultural data~\cite{ref19}. The role of big data analytics in Agricultural 4.0 involves transforming raw data into usable insights~\cite{ref18}. Machine learning algorithms are utilized to evaluate data to discern patterns, make predictions regarding crop diseases, optimize planting schedules, and effectively manage resources. According to~\cite{ref19}, this technology enables the implementation of data-driven decision-making processes, resulting in increased yields and resource conservation.

\subsection{Artificial Intelligence (AI)}

To be able to perform tasks traditionally effectively and efficiently that require human intelligence, such as learning from data, identifying patterns, and making informed decisions, artificial intelligence is the field and tool that uses such technologies to perform those tasks. Some of the primary AI models used in agriculture are image identification, natural language processing, and predictive modeling, according to Han et al. Artificial Intelligence plays a critical role in smart farming, and it is being utilized to handle a large number of datasets from all the sensors connected. These AI applications are designed to aid in the detection of crop diseases, monitor livestock health, and optimize farm operations. AlZubi et al.~\cite{ref14} found that AI models and algorithms in decision support systems have improved efficiency and production in major areas.

\subsection{Machine Learning (ML)}
Machine learning is another major technological field that uses algorithms-based models and data analytics to help automate operations, predict, and gain valuable insights from the datasets it has gathered~\cite{ref19}. Machine Learning is critical in intelligent farming and can be applied in agriculture to aid crop monitoring, disease detection, and decision-making processes. The study in~\cite{ref20} highlights that machine learning has the potential to improve automated processes and provide farmers with valuable insights, such as forecasting agricultural production, early disease identification in crops and livestock, and enhancing farm operational efficiency that will increase production.

\subsection{Robotic and Automation}

Another emerging technology which plays a vital role in smart farming is the field of Robotics. This is the advancement of machines that seek to operate independently or with minimal help of human interaction, such as autonomous tractors, harvesters, and drones used in agriculture and autonomous scarecrow~\cite{ref22}. Automations and Robotics plays a vital role in Agricultural 4.0 by focusing on the reduction of manual labor in farming by leveraging on the use of robotics and automation. Autonomous machines technology can perform activities like seed sowing, weed removal from the farm, and crop harvesting, while on the other hand drones can be used to conduct airborne surveys, detect issues, and counter interventions while enhancing labor productivity and reducing costs.

\subsection{Drones and Unmanned Aerial Vehicles (UAVs)}
Unmanned Aerial Vehicles (UAV) and Drones are mostly called autonomous aircraft equipped with cameras, sensors, and GPS technology. Which operates without a human pilot on board but is controlled from a different place; it is used to help collect data, monitor water, and create maps through data collection and analysis~\cite{ref21}. According to I. V. Kovalev et al.~\cite{kovalev2023conceptual}, drone UAVs play a crucial role in smart farming by monitoring crops to know the health status, livestock, and farm infrastructure and can also serve as security guards. These technological devices use high-resolution imaging to evaluate crop conditions, detect pests, and regulate irrigation practices. On the other hand, UAVs provide a cost-efficient method for gathering data across large agricultural regions.

\subsection{Smart Sensors}
Smart sensors are crucial in Agriculture 4.0, providing real-time data that helps farmers make informed decisions, optimize resource utilization, and improve farm management. The sensors are used to monitor environmental factors, which can create awareness of climate change, assess soil health, and enable precise irrigation~\cite{ref23}. They also aid in livestock management by tracking vital signs and movement patterns of the animals on the farm. Local weather data helps farmers prepare for extreme weather events and optimize resource allocation. With the help of the sensors, there is remote monitoring, which allows farmers to adjust equipment, irrigation systems, and climate control systems remotely. Data integration and analysis, often integrated with other sources, provide valuable insights for decision-making and resource management. Smart sensors also optimize energy usage by monitoring energy-intensive processes, reducing waste and operational costs~\cite{ref18}. They also ensure product traceability, ensuring the agricultural products are well kept in a secure and safe environment during transportation and storage.

The integration of technologies like IoT, AI, ML, robotics, and drones in agriculture has led to enhanced efficiency, sustainability, and productivity~\cite{ref2}. These technologies enable farmers to maximize resource use, minimize environmental impact, and improve food safety and traceability, thereby accelerating the progression toward Agricultural 4.0. Smart sensors are electronic devices with communication capabilities that collect and transfer data from the environment. They are primarily used in Agriculture 4.0 to monitor parameters like soil health, crop growth, and livestock well-being.

\section{Key Challenges and Concerns Associated with Agricaltural 4.0}

The emergence of Agricultural 4.0 presents significant advantages, although it also encounters various obstacles and concerns that require attention to ensure its sustainable and conscientious progression. Below are some primary issues and concerns that are closely linked to the implementation of Agricultural 4.0.

\subsection{Data Security and Privacy}

The rapid growth and exchange of extensive quantities of data in the context of Agricultural 4.0 give rise to notable concerns over data security and privacy. The repercussions of unauthorized access to sensitive farm data or the misuse of personal information might be severe~\cite{ref24}. Farmers and relevant parties must establish effective data security protocols to safeguard against cyber threats and breaches of data securit~\cite{ref24}. To ensure the protection of privacy, it is imperative to build robust systems for data ownership and permission.

\subsection{Digital Divide and Accessibility}
There exists a disparity in the accessibility of technologies and resources required for the implementation of Agricultural 4.0 among farmers. The digital gap, encompassing discrepancies in technology access and digital literacy, has the potential to worsen pre-existing inequalities~\cite{ref25} further. It is essential to make concerted efforts to address the digital gap, enabling small-scale and resource-constrained farmers to avail themselves of the advantages of Agricultural 4.0. The implementation of policies, the establishment of training programs, and the adoption of cost-effective technological solutions have the potential to mitigate this difficulty effectively.

\subsection{Environmental Sustainability}
The adoption of Agricultural 4.0 has the potential to enhance resource management. Yet, there exists a legitimate concern over the unintended consequences of rapid technological implementation, such as the overexploitation of resources, heightened energy consumption, and destruction of the environment Kalantari et al.~\cite{ref28}. The integration of sustainable practices and laws is crucial in the adoption of Agricultural 4.0. Achieving long-term sustainability necessitates a careful balance between technology-driven efficiency and environmental responsibility.

\subsection{Dependency on Technology}

The increasing reliance of agriculture on technological advancements raises concerns about its vulnerability to technical malfunctions, system breakdowns, or external disruptions~\cite{ref26}. Farmers must establish contingency plans and support structures to limit the dangers associated with dependence on technology. Implementing redundancy measures and backup solutions can effectively enhance the resilience of agricultural systems~\cite{ref10}. 

\subsection{Ethical and Social Implications}
The integration of advanced technology in agriculture raises ethical and social concerns, including job displacement, rural community implications, and animal treatment in smart livestock farming~\cite{ref27}. Ethical considerations are crucial in Agricultural 4.0 development, requiring collaboration between policymakers, farmers, and technology developers to optimize societal benefits while mitigating potential negative impacts. Addressing ethical concerns is essential for the responsible and sustainable advancement of Agricultural 4.0. Partnerships between government, researchers, industry stakeholders, and farmers are crucial for addressing complex challenges and utilizing technology to advance agriculture and society.

\section{Benefits of Agricultural 4.0}
Agriculture 4.0, driven by cutting-edge technologies and data-centric solutions, presents a wide range of advantages that possess the capacity to transform farming methodologies and the wider agricultural industry. These advantages not only tackle current obstacles but also provide the groundwork for a more sustainable and efficient future. 

\subsection{Increased Productivity}
Using real-time data obtained from sensors and devices empowers farmers to make well-informed decisions about irrigation, fertilization, and pest management. Consequently, this practice contributes to enhancing agricultural yields and livestock output. With the ongoing advancement of technology, it is anticipated that there will be more enhancements in productivity. Implementing advanced AI algorithms, robots, and automation systems is expected to yield significant improvements in efficiency within the agricultural sector~\cite{ref28}. This technological progress will empower farmers to increase their food production while simultaneously reducing resource consumption.

\subsection{Resource Optimization}
Agriculture 4.0 is a framework that advocates for the use of resource-efficient practices in farming. Through the real-time monitoring of soil conditions, weather patterns, and crop health, farmers can optimize the allocation of resources, such as water and fertilizers, in a more efficient manner~\cite{ref6}, the implementation of this approach leads to a decrease in waste generation and a reduction in the overall environmental footprint. The future trajectory of Agricultural 4.0 is the utilization of sophisticated sensors and analytics driven by artificial intelligence to attain enhanced levels of resource optimization. Predictive models possess the capability to foresee resource requirements, hence guaranteeing accurate allocation~\cite{ref28}.

\subsection{Sustainability}
Agricultural 4.0 offers a significant advantage in terms of sustainable farming practices and is essential in promoting environmental stewardship through the conservation of resources, reduction of chemical usage, and minimization of waste. The implementation of sustainable practices is of utmost importance in ensuring the attainment of long-term food security~\cite{ref6}. The forthcoming developments in sustainability will primarily concentrate on implementing carbon sequestration techniques, integrating renewable energy sources, and reducing the environmental impact associated with agricultural practices~\cite{ref29}.

\subsection{Food Safety and Traceability}
The importance of traceability within the contemporary food supply chain cannot be overstated. Agriculture 4.0 enables the comprehensive surveillance and oversight of every phase of the agricultural process, spanning from the cultivation stage to the final consumption stage~\cite{ref6}; this technology improves food safety, facilitates prompt recalls in the event of contamination, and satisfies customer expectations for transparency. Blockchain technology will probably assume a substantial role in guaranteeing traceability and food safety in the forthcoming years. The system provides safe and unchangeable documentation of food production and delivery~\cite{ref6, ref24}.

\subsection{Improved Livestock Management}

The benefits of Agricultural 4.0 are extended to the domain of livestock farming. IoT devices are utilized for the purpose of monitoring many aspects related to animal well-being, including health, behavior, and environmental circumstances~\cite{ref30}; the factors result in a decrease in disease outbreaks, improved feeding practices, and greater animal welfare. The ongoing advancements in technology are expected to further enhance the management of livestock, offering the possibility of autonomous feeding and treatments for maintaining their health.

\subsection{Market Access}
Agricultural 4.0 facilitates improved market accessibility for farmers through the provision of up-to-date information pertaining to product quality, quantity, and demand~\cite{ref27}; this phenomenon enables the promotion of economic expansion within rural regions, mitigates the occurrence of post-harvest losses, and enhances the financial gains experienced by farmers. The future will prioritize the expansion of access to global markets and the promotion of fair trade. The ongoing advancement of technology is expected to further narrow the disparity between rural and urban regions.
Future iterations of Agricultural 4.0 may include 5G technology for improved connectivity, quantum computing to aid in complex data analysis, and the widespread adoption of sustainable and regenerative practices to combat global issues like climate change~\cite{ref6, ref31}.

\section{Related Work}

These case studies highlight the concrete and transformational impacts of Agricultural 4.0 technology on numerous aspects of farming, ranging from crop cultivation and water management to livestock health and new reforestation approaches. Agricultural 4.0 technologies have the potential to revolutionize farming in many ways.

\subsection{Precision Agriculture: A Success Story}

In~\cite{ref32}, the authors describe Precision agriculture as a prime example of how Agricultural 4.0 technologies have fundamentally transformed traditional agricultural methods. This involves using IoT sensors and GPS technology to create precise field maps for agricultural practitioners, identifying soil composition, moisture levels, and crop vitality disparities. Providing comprehensive maps enables farmers to tailor their agricultural techniques to the unique characteristics of each section within their farms.

In~\cite{ref33}, the authors propose that using IoT soil sensors and GPS-guided tractors in agriculture has enabled farmers to collect real-time data on soil conditions, enabling better management of irrigation and fertilizing programs. This has reduced overlap and resource wastage, while historical data analysis allows farmers to predict disease outbreaks and implement preemptive measures, thereby mitigating crop losses. Overall, these advancements have significant benefits for agriculture.

\subsection{Smart Irrigation Systems: Water Management Excellence}

Water scarcity is a pressing global concern, and agriculture is a major consumer of this finite resource. Intelligent irrigation systems represent a solution by optimizing water usage through IoT sensors and real-time data analysis [34]. Smart irrigation systems using IoT and sensory systems are crucial for achieving the United Nations' Sustainable Development Goals, particularly Goal 6 and Target 6.4. These systems conserve water, reduce usage, and improve crop understanding, but they also present challenges like crop wastage and environmental impact~\cite{ref35}.

Water management is crucial in water-scarce countries, particularly in agriculture. Low-cost sensor-based irrigation systems using IoT technologies are gaining interest and monitoring water quantity, quality, soil characteristics, and weather conditions, with communication technologies playing a crucial role.[34] and there is a need for further research in smart irrigation systems to address challenges and implement best practices while also demonstrating the reduction of energy expenses~\cite{ref36}. The irrigation system utilized weather forecasts and historical data to schedule irrigation events, reducing the risk of excessive watering.

\subsection{Vertical Farming: Revolutionizing Crop Production}

Vertical farming is a new agricultural technique that involves growing crops in vertically arranged tiers, often within enclosed environments, relying on technological advancements for efficiency [37]. Vertical farming is a sustainable approach to urban food provision, addressing food security, population growth, farmland shortages, and greenhouse gas emissions. A study by~\cite{ref28} indicates that Vertical farming involves:

\begin{itemize}
	\item Cultivating plants with livestock on vertically inclined surfaces.
	\item Increasing food production.
	\item Maintaining quality and safety.
	\item Providing educational facilities.
\end{itemize}

It can be implemented anywhere in the world and address issues related to the use of fecal matter as fertilizer. Locally produced, fresh, and organic food can also improve overall health. However, some challenges of implementing Vertical farming include economic feasibility, codes, regulations, and a need for more expertise~\cite{ref37}. Using a controlled environment has significantly reduced the need for pesticides, resulting in healthier and more visually appealing produce.

\subsection{Livestock Monitoring: Ensuring Animal Welfare}
Agricultural 4.0 technology offers benefits beyond crop farming but also enhances animal welfare and productivity in livestock farming~\cite{ref38}.

Precision agriculture in livestock farming uses smart technologies to monitor animal performance and manage field variability. Precision Livestock Farming systems automate the collection, analysis, and use of feeding-related information, thereby enhancing animal productivity, farm profitability, product quality, welfare, and environmental sustainability~\cite{ref30}. Precision Livestock Farming technologies optimize resource use efficiency and animal productivity by electronically measuring production components, interpreting information, and controlling processes~\cite{ref30}. However, further research and collaboration with private companies are needed for the full implementation of Precision Livestock Farming systems. GPS monitoring technology improved cow allocation to productive pasture areas, leading to increased milk output and overall herd health improvement.

\subsection{Agri-Tech Startups: Pioneering Innovations}

The Agricultural 4.0 era has spurred numerous startups focusing on developing advanced technologies specifically for the agricultural sector~\cite{ref21}. Drones are increasingly used in agribusiness and precision farming, providing aerial photography for vegetation analysis, density, germination, growth, and land productivity prediction. They also help identify oppressed vegetation. Traditional satellite remote sensing is used for NDVI calculations, but affordable unmanned aerial vehicles reduce costs and increase efficiency. IoT technology is transforming agriculture by enabling farmers to monitor and manage various aspects of farming through smart devices and sensors. Drones also facilitate crop health surveillance and timely action against pests or diseases~\cite{ref15}.

\section{Proposed Framework}
\begin{figure}[t]
	\centerline{\includegraphics[width=8.5cm, height= 6 cm]{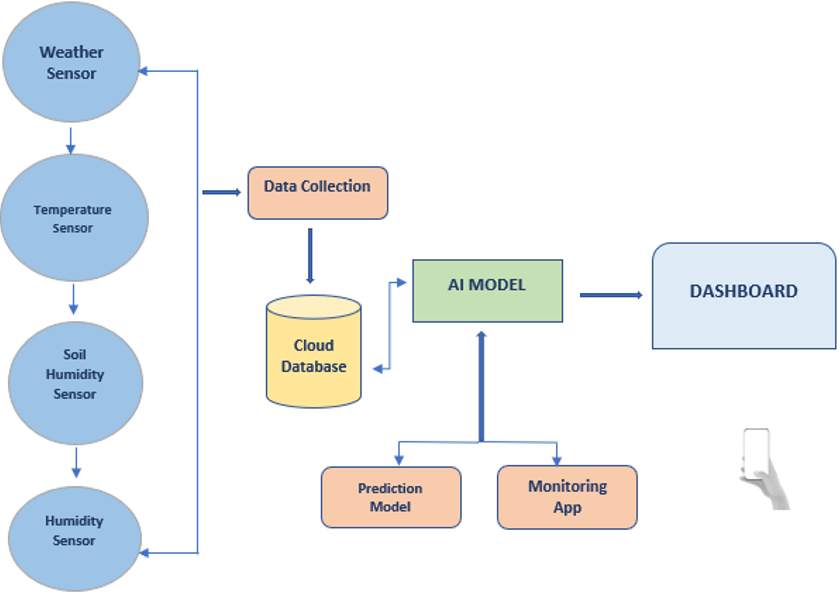}}
	\caption{The proposed system and dashboard services for smart farming framework.}
	\label{fig4}
\end{figure}

Agricultural 4.0 is a crucial aspect of modern agriculture, allowing farmers to optimize crop yields and help reduce resource usage. This proposed framework proposes a system that will integrate weather, temperature, soil humidity, and humidity sensors to improve farm management and sustainability. The proposed system and dashboard services for smart farming is shown in Figure~\ref{fig4}.

The weather sensors will provide real-time data, allowing the farmers to make informed decisions about planting, harvesting, irrigation, pest and disease management, and climate control. The temperature sensors will aid in the optimization of planting times, monitor pest and disease conditions, and provide fertilization recommendations.
The humidity sensors optimize ventilation and predict disease outbreaks. This data-driven approach will strategically enhance crop yield and quality, promoting sustainability and productivity in the agriculture industry.

The smart farming system that has been proposed represents a comprehensive strategy aimed at modernizing the agricultural sector, improving efficiency and output, and fostering sustainability in response to worldwide issues. Through the utilization of cutting-edge technology, sophisticated data analytics, and autonomous machinery, farmers have the ability to enhance the efficiency of resource allocation, mitigate detrimental effects on the environment, and effectively respond to the challenges posed by shifting climate conditions. The increasing integration of technology and data in agriculture is anticipated to have a significant impact on food security and environmental conservation, with smart farming positioned as a crucial component in this transition.

\subsection{Opportunities of Agricultural 4.0}
Agricultural 4.0 technologies optimize resource allocation, increase crop yields, and enhance livestock management, increasing profitability and economic stability. They use IoT sensors and real-time data to promote sustainable farming practices, reduce costs, and improve market access. These solutions reduce environmental impact, meet consumer demand for ethical food, and improve animal health and welfare. They foster innovation and economic growth, benefiting technology companies and startups~\cite{ref6, ref10}.

\subsection{Challenges of Agricultural 4.0}

Liu et al.~\cite{ref10} highlight the potential of agricultural 4.0 technologies in transforming the agriculture sector but also highlight challenges such as high initial investment and the digital divide. Small-scale farmers may need help to accept these technologies due to their unpreparedness for financing options or subsidies. Addressing these issues is crucial for promoting equal access to technology and digital literacy.
Smart farming faces data privacy and security challenges, with many farmers lacking equal access to technology and literacy. Robust measures and education on data privacy are crucial, while regulatory compliance and ethical considerations like fair labor practices and responsible data use are also crucial~\cite{ref6}. To fully explore the potential of smart farming, Stakeholders must be trained to effectively utilize smart farming technologies, ensure fair trade practices, conduct environmental impact assessments, foster collaboration, and data sharing, and address the social and community impacts of technological advancements in agriculture to fully realize their potential~\cite{ref12}.

\subsection{Limitations}

Agricultural 4.0 technologies face significant challenges due to substantial financial and resource investments, particularly for small-scale farmers who struggle to afford or access essential technical gear like sensors, drones, and automated machinery. Data privacy and security are crucial in farming due to data-driven technologies, posing cyber risks, unauthorized access, and potential abuse, necessitating literature study to address these issues~\cite{ref6, ref31}. The lack of legislative frameworks and ethical considerations for data ownership and sharing in the agriculture 4.0 ecosystem could hinder technology adoption and effectiveness. Modern farming relies heavily on technological devices, increasing the risk of system failures, cybersecurity breaches, and technological obsolescence, which can disrupt agricultural operations, particularly for farmers lacking resources or expertise.

\subsection{Future of Agricultural 4.0}
Agricultural 4.0 is a set of technologies that will revolutionize the agriculture sector by enhancing efficiency, productivity, and sustainability~\cite{ref13}. The proposed framework aims to empower farmers by enabling them to monitor activities on the farm using sensors that track soil conditions, temperature, and humidity. Securing the collected data is crucial, considering that data security is a significant concern in smart farming. The AI model utilizes data-driven technologies, such as machine learning (ML), to conduct predictive analysis based on the gathered data, assisting farmers in making informed decisions. Emphasizing usability in technology development, the dashboard is designed to help farmers navigate and monitor the system seamlessly. 5G networks offer rural connectivity, IoT data transmission, AI-driven decision-making, and blockchain technology for food safety, enhancing farmers' efficiency and consumer visibility~\cite{ref10}. Regenerative farming practices are becoming increasingly important, with Agricultural 4.0 technologies playing a crucial role in implementing and monitoring these practices. Advances in biotechnology and genetic engineering will lead to improved disease resistance, higher yields, and better nutritional profiles~\cite{ref7, ref27, ref29}.

Robotics and autonomous farming systems will enhance tasks like planting, harvesting, weed control, and pest control, while integrating climate change adaptation strategies for optimal resource use. Agricultural 4.0 will improve farming efficiency and productivity and contribute to sustainable and resilient agriculture, addressing challenges posed by population growth, climate change, and resource constraints.

\section{Conclusion}
Agriculture 4.0 signifies an evolution in the agricultural industry, driven by technological advancements and the adoption of data-centric decision-making processes. The use of the Internet of Things (IoT), big data analytics, artificial intelligence (AI), Machine Learning, robotics, and automation has led to enhanced efficiency, sustainability, and productivity in the field of agriculture. Case studies have provided evidence of the effective utilization of these technologies, illustrating their practical influence on agricultural productivity, efficient allocation of resources, and the well-being of animals. Nevertheless, it is significant to acknowledge and confront certain obstacles to guarantee the appropriate implementation of Agricultural 4.0, including but not limited to data security issues, infrastructure limitations, and ethical considerations. Quantum computing could significantly enhance smart agriculture. 

To effectively harness the capabilities of Agricultural 4.0, the paper emphasizes the importance of implementing Agricultural 4.0 through ethical considerations. It focuses on training, capacity building, funding, collaborative networks, and a farmer-centric approach to ensure a sustainable future for global farming communities, ensuring a prosperous future.

\bibliographystyle{ieeetr}
\bibliography{Kojo_references}

\end{document}